\documentclass[prb,showpacs,twocolumn,superscriptaddress, a4paper]{revtex4}
\usepackage{graphicx}
\usepackage{amsmath}
\usepackage{bm}
\usepackage{color}
\usepackage{hyperref}
\usepackage{natbib}
\begin{document}
\title{Hatano-Nelson model with a periodic potential}

\author{F.~H\'ebert}
\affiliation{INLN, Universit\'e de Nice-Sophia Antipolis, CNRS; 
1361 route des Lucioles, 06560 Valbonne, France}
\author{M.~Schram}
\affiliation{Physics Department, Massachusetts Institute of Technology,
 77 Massachusetts Avenue Cambridge, MA 02139-4307, USA}
\author{R.T. Scalettar}
\affiliation{Physics Department, University of California, Davis, California 95616, USA}
\author{W.B. Chen}
\affiliation{School of Mathematical Sciences, Fudan University, Shanghai, 200433, P.~R.~China}
\author{Z. Bai}
\affiliation{Departments of Mathematics and Computer Science, University of California,
Davis, California 95616, USA}

\date{\today}

\begin{abstract}
We study a generalisation of the Hatano-Nelson Hamiltonian
in which a periodic modulation of the site energies is present in
addition to the usual random distribution. The system can then become
localized by disorder or develop a band gap, and the eigenspectrum shows
a wide variety of topologies.  We determine the phase diagram, and
perform a finite size scaling analysis of the localization transition.
\end{abstract}


\maketitle

\section{Introduction}

The Anderson transition\cite{anderson58,thouless74}, in which
eigenstates of non-interacting electrons are localized by disorder, has
long been a paradigm of a metal-insulator transition in solid state
physics.  The metallic phase  occurs only in three dimensions, since any amount
of randomness localizes all particles in one or two dimensions. In three
dimensions, a mobility edge separates localized states whose energies
lie at the edges of the spectrum, from extended states whose energies
lie in the middle.  The mobility edges appear for a
critical value of the disorder and move with the disorder until no
extended states remain. If the Fermi energy is located below or above
the mobility edge, inside the localized states, the system is then an
insulator.  Recently, research on this subject has experienced renewed
interest as localization of cold atoms in the presence of a disordered
optical potential has been observed \cite{aspect08, inguscio08}. Whereas
experiments in solid state physics are intrinsically limited as
interactions between electrons cannot be neglected, cold atoms allow the
exploration of the non-interacting regimes with possibilities of control
that could not be achieved before.

Another, simpler origin for an insulating behavior in a solid is the
presence of gaps due to a periodic potential applied to the particles.
These two possible insulating phases are in competition with each other
as they are based on opposite behaviors.  The Anderson insulator is due
to the localized nature of the wave function, whereas the gapped
insulator is due to its extended nature.  A third source of insulating
behavior is interactions between particles.  In many ways this is the
most difficult situation, since it involves treating the many-body
problem accurately.  In the systems with both interactions and
randomness, competition between Mott and band insulating phases have
been shown to result in metallic phases, as the two sources of
insulating behavior  counteract each other \cite{paris07,bouadim07}.
The possibility of metallic phases arising due to the addition of
interactions to two dimensional disordered systems has also been an area
of great interest \cite{abrahams79,lee85,belitz94}.

The goal of this article is to study the interplay between the two
sources of single particle insulating behavior- disorder and gaps
due to a periodic potential- in the context of non-Hermitian
matrices.  For this we
will use a modified version of the Hatano-Nelson model (HNM), that
includes at the same time a periodic potential that generates gaps in the
eigenspectrum and a disordered potential that will drive a localization
transition.  The HNM \cite{hatano96, hatano97} is a single particle,
random, non-Hermitian Hamiltonian that was introduced to study the
motion of magnetic flux lines in disordered type II superconductors, the
path integral representation of the HNM being analogous to a 
classical model used to describe the flux lines. The HNM shows a
localization transition in one
dimension\cite{hatano96,hatano97,hatano98, feinberg99}. It is then a
very convenient tool to study the localization transition numerically,
as it allows the study of systems with large linear sizes. 

The paper is organised as follows. In
the Sec.~II we introduce the HNM and recall some basic properties.
In Sec.~III, we present the different regimes which can be
observed.  Sec.~IV contains a detailed analysis of some
of the transitions appearing in the generalized HNM.
The conclusion describes some of the connections the additional term we
introduce into the Hatano-Nelson model has with the eigenspectrum of
several physical systems for whose properties random matrices have
been used.

\section{Modified Hatano-Nelson model}

\subsection{Hatano-Nelson model}
The Hatano-Nelson model is a discrete model similar to the Anderson
model used to study localization. In one dimension,
\begin{equation}
\mathcal{H}_{\rm HN} =\sum_{x=1}^L  \left[-\frac{t}{2}
\left( e^h c^\dagger_{x+1} c^{\phantom{\dagger}}_x 
+ e^{-h} c^\dagger_{x} c^{\phantom{\dagger}}_{x+1} \right)
 + \mu_x n_x  \right]
\label{eq:ham}
\end{equation}
Here $c^\dagger_x (c^{\phantom{\dagger}}_x)$ are the usual fermionic
creation (destruction) operators and $n_x = c^\dagger_x
c^{\phantom{\dagger}}_x$ is the number of particles on site $x$.  The
first term of the Hamiltonian describes the hopping of particles between
sites, and the parameter $h$ controls the asymmetry between the hopping
amplitudes in the left and right directions.  We will choose the
random sites energies
$\mu_x$ according to  a uniform distribution on the interval
$[-\Delta / 2, +\Delta /2]$. The hopping parameter $t$ is set to one to
fix the energy scale and $L$ is the number of lattice sites.

\begin{figure}[h]
\includegraphics[width=0.5\textwidth]{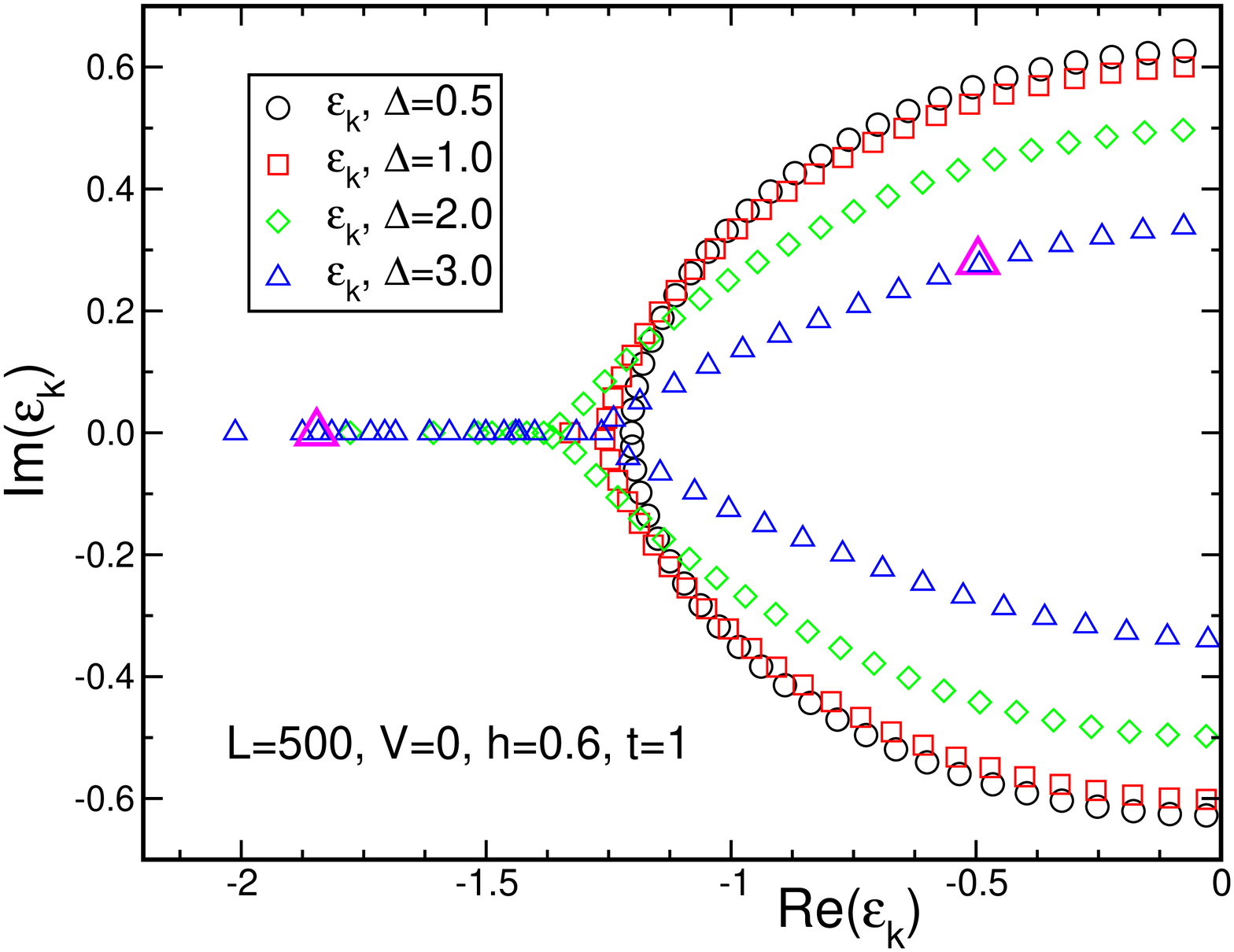}\\
\caption{(Color online) Typical spectra of the Nelson-Hatano model.
In the presence
of disorder, localized states (real eigenvalues)
appear for $\Delta$ larger than a critical value. 
Half of the spectra are shown and the spectra are approximately 
symmetric around the imaginary axis. The two larger up triangles
correspond to the eigenstates used for Fig.~\ref{densities}.
\label{typical_V0}}
\end{figure}

In the absence of disorder ($\Delta=0$) the eigenenergies are
given by,
\begin{equation} 
\epsilon_k = \frac{t}{2}\left( e^{h+ik \cdot2\pi/L} +
e^{-h-ik \cdot 2\pi/L}\right)
\label{eq:deq0spectrum}
\end{equation} 
where $k = 1, 2\cdots L$.
All the eigenvalues (except $k=L/2,L$) 
are complex and lie on an ellipse in the complex
plane, centered at the origin.  When disorder is turned on, $\Delta \ne
0$, the eigenvalues of the $d=1$ HNM remain confined to one dimensional
curves in the complex plane \cite{feinberg99b}, however, some of the
eigenvalues become real.  The corresponding eigenstates become localized
(Fig.~\ref{typical_V0}). 
We will continue to use the notation $\epsilon_k$ for the
eigenvalues for nonzero $\Delta$ even though momentum is no longer
a good quantum number due to the breaking of translation
invariance by the disorder.  $k$ will be understood to be a generic
mode label.

One way to understand the appearance of real eigenvalues is to consider
a system with open boundary conditions (OBC)\cite{hatano96, hatano97}.
In this case, all the eigenvalues of $\mathcal{H}_{\rm HN}$ are real,
despite the fact that the Hamiltonian is not Hermitian.  In fact, the
eigenspectrum is independent of $h$, since the $e^{\pm h}$ factors can
be ``gauged away" by a suitable redefinition $\tilde \phi(x) = e^{-hx}
\phi(x)$, of the eigenvector components.  The HNM spectrum becomes
identical to that of the Anderson model.  On a system with periodic boundary
conditions (PBC), however, this eigenvector transformation is no longer
possible, and some eigenvalues are complex (see Fig.~\ref{typical_V0}).

The dramatic difference in the spectrum between OBC and PBC
of course reflects a rather general property of the localization
transition: A localized state is insensitive to the boundary conditions
and its eigenvalue should remain real, even in the presence of PBC. On
the contrary, an extended state, for example a plane wave state, is
sensitive to the boundary conditions and its eigenvalue should be
complex in the PBC case (Fig.~\ref{typical_V0}).

Le Doussal\cite{ledoussalunp} has further clarified the relation
between the eigenvector tranformation 
and the localization length by pointing out that the eigenvalues will
remain real if the localization decay $e^{-l/\xi}$ is more rapid than
the growth $e^{hl}$ induced by the transformation.  As a consequence,
the localization length $\xi$ does not
diverge at the transition, but takes on the value $\xi=1/h$
at the bifurcation point in Fig.~\ref{typical_V0} where the
line of real eigenvalues splits off into the complex plane. 

For the particular case of a Cauchy distribution of
randomness, $P(\Delta)=\gamma/\pi(\gamma^2+\Delta^2)$ the location
of the bifurcation point is known \cite{hatano98} to be given by
${\rm Re} [\epsilon_k] = \sqrt{(2t\,{\rm sinh}\,h)^2 - \gamma^2}
\, /\, {\rm tanh} \, h$, and
${\rm Re}[\epsilon_k]=0$ at the critical value 
$\gamma_c = 2 t {\rm sinh} h$.  At this point the randomness is
sufficiently large that all eigenvalues are real, and all
modes are localized.

\begin{figure}[h]
\includegraphics[width=0.5\textwidth]{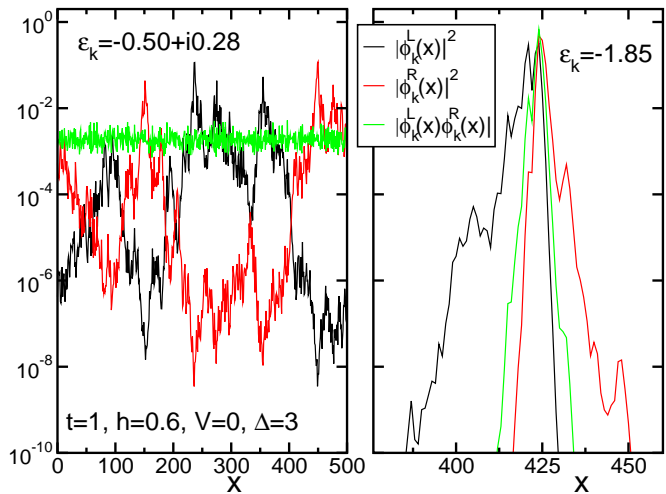}\\
\caption{(Color online) Three different possible definitions of the density in an
extended (left) and localized (right) states corresponding to 
different eigenvalues $\epsilon_k$ of the same system. In general
$|\phi_{k}^{L,R}(x)|^2$
show quite strong fluctuations, thus making $\left|\phi_{k}^L(x)
\phi_{k}^R(x)\right|$ a better quantity to identify the different states.  \label{densities}}
\end{figure}

More direct evidence of the localized nature of the eigenstates with
real eigenvalues can be obtained by directly looking at how the
corresponding density varies with distance\cite{hatano98}.  For example,
we can study the participation ratio $P_k$ for a given eigenvalue
$\epsilon_k$. As the Hatano-Nelson model is non-Hermitian, the left
eigenvector $\langle k | = \sum_x \langle x | \phi_{k}^L(x) $ and right
eigenvector $|k\rangle  = \sum_x \phi_{k}^R(x) |x\rangle$ are not
Hermitian conjugates.  It was shown\cite{hatano98} that the probability
density in these systems is given by $\left|\phi_{k}^L(x)
\phi_{k}^R(x)\right|$ which is a better behaved quantity
than $|\phi_{k}^L(x)|^2$ or $|\phi_{k}^R(x)|^2$ 
although they give quite similar results for localized states (Fig.~\ref{densities}).

\begin{figure}[h]
\includegraphics[width=0.5\textwidth]{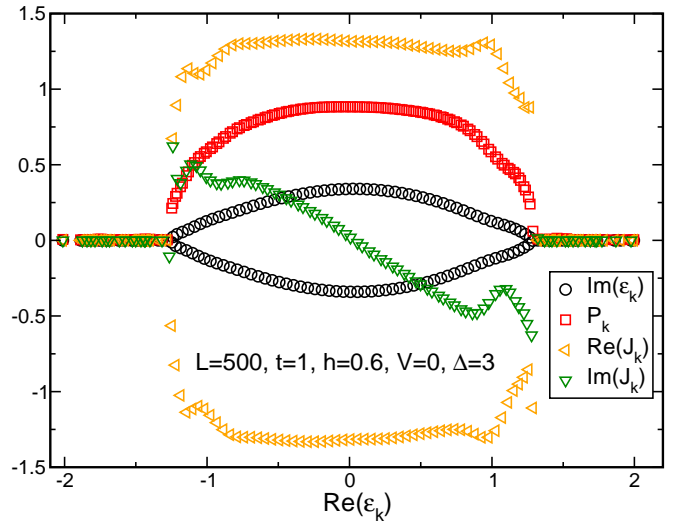}\\
\caption{(Color online) Different quantities which distinguish the extended from the localised
states in the Hatano Nelson model: the imaginary part of the eigenvalues
${\rm Im}(\epsilon_k)$, participation ratio $P_k$, 
and real ${\rm Re}(J_k)$ and imaginary ${\rm Im}(J_k)$ parts of the current.
All these quantities give consistent information
concerning which modes are localized or not.\label{quantities}}
\end{figure}

Then the participation ratio $P_k$ can be
defined as,
\begin{equation}
P_k = \frac{1}{L}\sum_x \left|\phi_{k}^L(x) \phi_{k}^R(x)\right|^2
\label{eq:participation}
\end{equation}
$P_k$ corresponds to the portion of the lattice sites that is occupied
for a given eigenvalue. It goes to zero for a localized state
in a large system and goes to a number of
order unity for an extended state when
the density distribution is spread
equally over all the sites. 
Fig.~\ref{quantities}
shows that the vanishing of the participation ratio aligns 
nicely with the energy
at which the eigenvalues are real. Similar participation ratios can be
defined with other definitions of the density $P^{L,R}_{k} = \sum_x
|\phi_{k}^{L,R}(x)|^2/L$. But, as for the densities
(Fig.~\ref{densities}) themselves, $P_k$ is better behaved than
$P^{L,R}_k$.

The current $J_k$ defined as $J_k = -i \partial \epsilon_k / \partial h$ \cite{hatano97} is
compared with the participation ratio in Fig.~\ref{quantities}. It gives similar
results for the extended or localized nature of the states, but is more difficult to calculate as it requires a numerical 
derivative (and then two diagonalization instead of one for the participation ratios) and
is difficult to define close to the mobility edges.

\begin{figure}[h]
\includegraphics[width=0.5\textwidth]{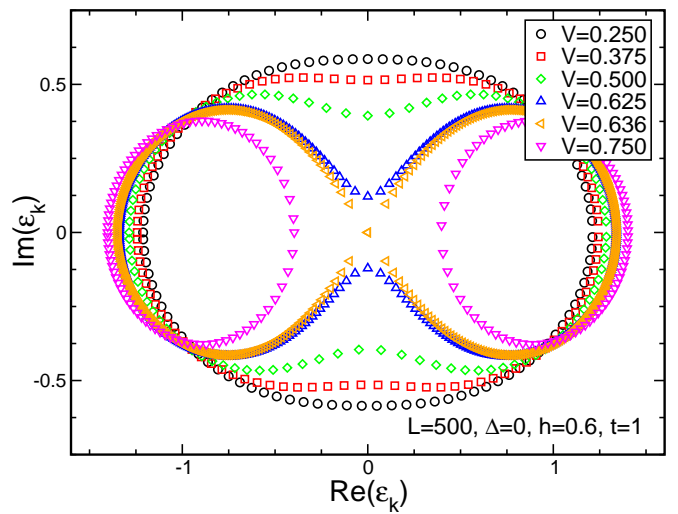}\\
\caption{(Color online) With the periodic potential, a gap is opened for 
$V > V^{\rm o}_{\rm c} = t \sinh(h)$ although all the states remain extended.
$V^{\rm o}_{\rm c} \simeq 0.636$ in the case shown, $t=1.0, h=0.6$.\label{typical_D0}}
\end{figure}

\subsection{Effect of a periodic potential}

Our interest here is in adding
a simple potential of period 2 to the original HNM,
\begin{equation}
\mathcal{H} = \mathcal{H}_{\rm HN} + \sum_{x=1}^L V \cos(\pi x)  n_x
\end{equation}
When $h=\Delta=0$, this term
generates a spectrum with two bands separated by a gap of
width $2V$.
In the non disordered case where 
$h \ne 0$ and $\Delta = 0$ we obtain the following eigenvalues
\begin{equation}
\epsilon_k = \pm \frac{t}{2}\sqrt{4V^2/t^2 + 2 + e^{2h+ i k \cdot 2 \pi/L}
+ e^{-2h -i k \cdot 2 \pi/L}}
\end{equation}
with $k = 1, 2\cdots L/2$. In this ordered case, the spectrum has a gap if
$V > V^{\rm o}_{\rm c} = t \sinh(h)$ and its width $W_{\rm G}$
is given by
\begin{equation}
W_{\rm G} = 2\sqrt{V^2 - t^2 \sinh^2(h)}
\end{equation}
Even when $\Delta = 0$, there is a transition 
in this model corresponding to the opening of the gap at 
half-filling (see Fig.~\ref{typical_D0}).
The gap opens with exponent 
$\beta=1/2$ since near
the transition 
$W_{\rm G} \simeq \sqrt{2V^{\rm o}_{\rm c}(V-V^{\rm o}_{\rm c})}$.

\section{Phase diagram}

In the presence of both disorder and the periodic potential,
the spectrum of the model can show several different topologies.
We calculate the eigenspectrum for one disorder realisation of a large size system ($L=1000$)
and analyse
the different topologies appearing.
We checked that these topologies were still
present for other sizes ($L=500, L=2000$) and other disorder realisations.

\begin{figure}
\includegraphics[width=0.5\textwidth]{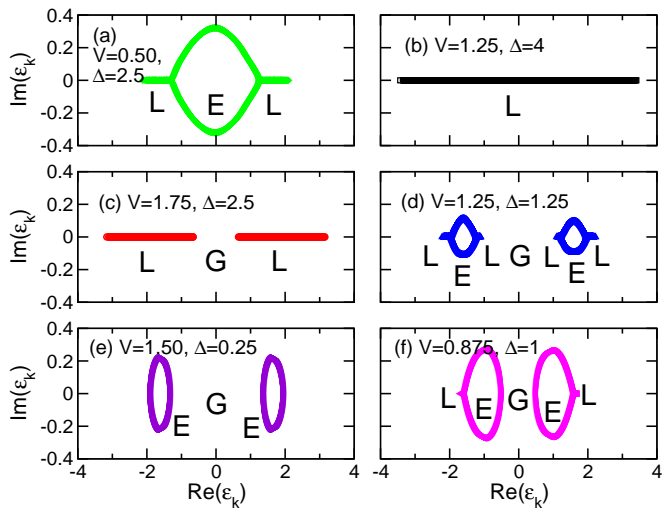}
\caption{(Color online) Different topologies of the eigenspectrum
for one realization of a $L=1000$ system and for $h=0.6$. 
The simple E topology (only extended states without a gap) and
the rarely observed LELEL are not shown.
See text for discussion.
\label{difftypes}}
\end{figure}

\begin{figure}
\includegraphics[width=0.5\textwidth]{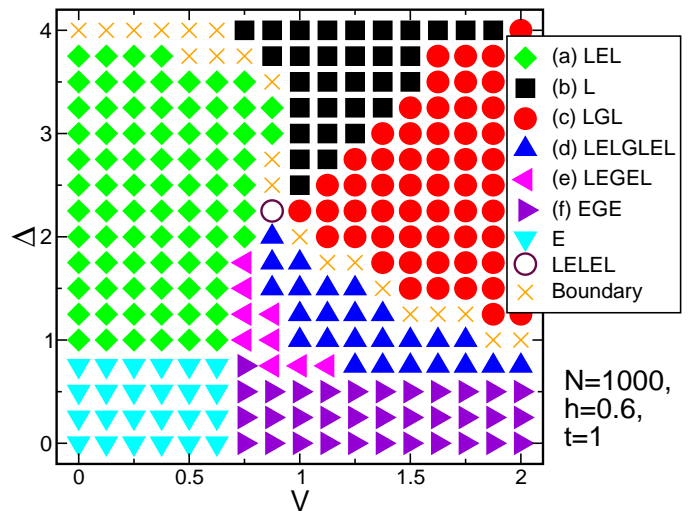}
\caption{(Color online) The different phases observed in the Hatano-Nelson model for a given realization
on a $L=1000$ system size and for $t=1, h=0.6$. At the boundary
between different phases, the behavior is sometimes irregular
for a finite size system
and does not fall into one of the identified categories. The labels (a)-(f)
correspond to the cases shown in Fig. \ref{difftypes}.
\label{phasediagram}}
\end{figure}

Scanning horizontally along the real axis in the complex eigenvalue
plane, we observe three types of regions in the spectrum. First we
observe cases of real eigenvalues corresponding to localized (L)
states. Second we observe regions with complex eigenvalues corresponding
to extended (E) states. Finally we observe some gaps (G) in the
spectrum. A given topology is characterized by the succession of these
different regions as one increases the real part of the eigenvalue, ${\rm
Re}(\epsilon_k)$.  The different topologies we observed are shown in
Fig.~\ref{difftypes} and their position in the $(V,\Delta)$ plane for a
fixed value $h=0.6$ is shown in Fig.~\ref{phasediagram}.  As an example,
the phase shown in Fig.~\ref{difftypes}(a) is denoted as the LEL phase
since we observe a central region of extended states (E) between  
two regions of localized states (L) with no gaps.

The location of these different phases  for $h=0.6$ is shown in Fig.~\ref{phasediagram}
for a given realization.
As expected, a gap appears for $V$ large enough. For moderate disorder
($\Delta < 2$), the opening of the gap happens approximately at the same
value $V^{\rm o}_{\rm c}={\rm sinh}\,0.6=0.636$ 
as in the ordered, $\Delta=0$ case.  Upon increasing
$\Delta$, the original low disorder E and EGE phases are modified as
some of the extended states are transformed into localized ones at the
edges of the extended regions, leading to the LEL and LELGLEL (or LEGEL)
phases respectively. In the LEGEL phase, localization first appears on
the external edge of the extended regions and not on the border of the
gap.  This is consistent with the fact that localized states
appear first for the extremal eigenvalues in the Anderson model.

In a very small region of the phase space, we also observe a LELEL
phase where the region between the
two extended blobs is filled by localized states but no gap
appears.  As $\Delta$ is
increased further, all the extended states disappear leading to the
large disorder L and LGL phases.  When $V$ and $\Delta$ are both large,
the transition from the L phase to  the LGL  phase occurs at $V_{\rm c}
= \Delta/2 $ which is the result expected in the one site limit, when
the hopping terms are neglected. In the one site limit the gap opens
like $W_{\rm G} \simeq 2(V-V_{\rm c})$ which is a different behavior
from the opening of the gap due to the periodic potential.

\begin{figure}
\includegraphics[width=0.5\textwidth]{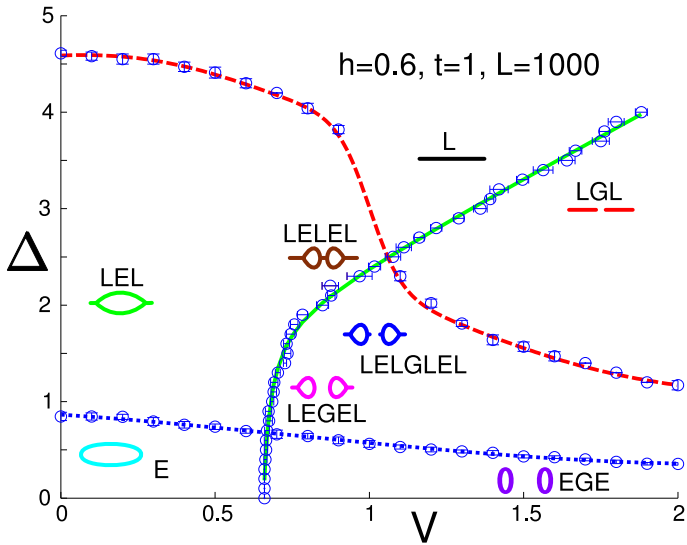}
\caption{(Color online) The phase diagram obtained by averaging several disorder realisations 
to calculate
the boundary between different regions.  The smooth lines are fits to the data.
The continuous green line marks the limit between gapped
and ungapped regions. The dotted blue, the limit between a spectrum without any
real values and a spectrum with real and complex values. The dashed red line is the limit
between a spectrum with real and complex eigenvalues and a spectrum
with only real values.  \label{phasediagram2}}
\end{figure}

For a given realisation, close to the boundary
between two phases, the spectrum becomes disordered 
and it is difficult to classify it as 
belonging to one of the simple topologies identified before.
These regions are indicated by ``$\times$" 
in the chosen example (Fig.~\ref{phasediagram}). To calculate
more precisely the boundaries between the different phases, 
it is then necessary to use different quantitative indicators 
and to average those over several disorder realisations.
First we measure the width of the gap
in the system and determine whether it is present or  not. 
Second we determine if the spectrum is made of
complex values (extended states) only, of a mixture of complex and real values
(extended and localized states), or of real values only.
We consider a $L=1000$ system and average over 10 realisations.
Drawing these three lines in the phase space, we obtain the
phase diagram shown in Fig.~\ref{phasediagram2}. With these
quantities, we cannot distinguish the LELGLEL from the
LEGEL phases, as well as the LELEL from the LEL phase.

\begin{figure}
\includegraphics[width=0.5\textwidth]{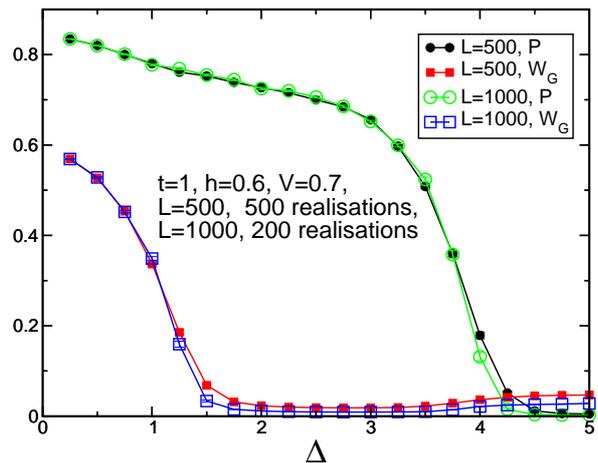}
\caption{(Color online) For $V=0.7$, upon increasing the disorder, the gap $W_G$ is closed
before the states becomes localised, which is shown by the participation ratio $P$ becoming zero.
Here, the quantities are averaged over only 5\% of the total 
number of eigenstates located in the middle of the spectrum,
to get rid of possible bias induced by states located at the edges
of the spectrum.  The measurements show the crossings of the two boundaries
of the phase diagram at largest $\Delta$ (gap closing and
all extended states vanishing) for $V=0.7$ in 
Fig.~\ref{phasediagram2}, but not the crossing of the boundary at
smallest $\Delta$ when localized states first appear.
\label{reentrant}}
\end{figure}

We observe an interesting reentrant behavior. If the system is in the 
EGE phase but the gap is not too large, when we increase $\Delta$,
the gap will be closed by the disorder but the states in the middle
of the spectrum will remain extended (Fig.~\ref{reentrant}).  Then we observe a
transition where increasing disorder drives the gapped system
into an ungapped phase. This can be understood as follows: as the disorder
is increased, the E regions have a tendency to flatten out as 
they evolves towards L regions with only real values. By doing so,
the E regions spread along the real axis. If the gap is not
to large, two gapped E regions can then merge in the process, which
is happening here.

\section{Transitions}

\subsection{Opening of the gap}

\begin{figure}
\includegraphics[width=0.5\textwidth]{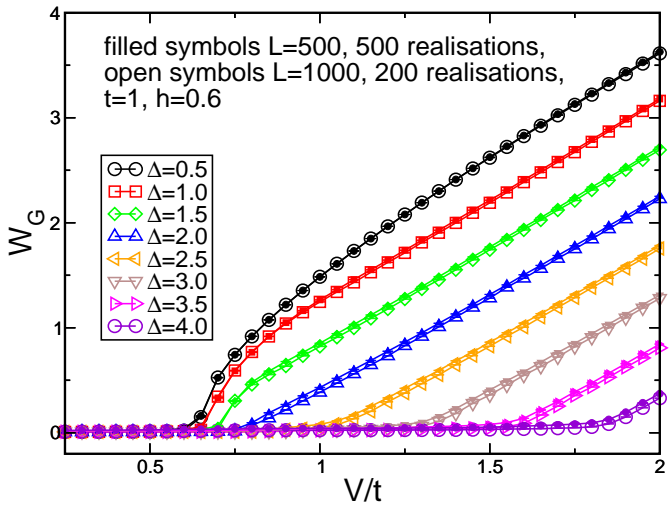}
\caption{(Color online) Evolution of the gap as a function of $V$. For $V< 2.0$, the gap
opens like $W_{\rm G} \propto \sqrt{V-V_{\rm c}}$ 
as in the non disordered case. For $V \ge 2.0$, the
opening of the gap is linear as expected for a transition in the disorder regime.
\label{gapopening}}
\end{figure}

As expected, we observe two different behaviors for the opening of the
gap for a fixed value of $\Delta$ as $V$ is increased (see Fig.~\ref{gapopening}). 
If $\delta = V-V_{\rm c}$, we
have a behavior $W_{\rm G} \propto \sqrt{\delta}$ as in
the non interacting regime, whereas 
$W_{\rm G} \propto \delta$ in the strongly disordered regime
Fig.~\ref{gapopening}. The transition between those two different behaviors
happens precisely at the point where, in the phase diagram Fig.~\ref{phasediagram2},
we are passing from the low disorder into the large disorder regime.
For $\Delta < 2$, the transition almost takes place at the non-interacting
critical value $V_{\rm c}^{\rm o}$, the transition occurring at nearly constant 
$V$ and with the square root behavior. 
Above this value, the transition takes place approximately
at $V_{\rm c} = \Delta /2$ with the linear behavior, 
which is characteristic of the disordered regime.

\subsection{Finite size analysis of the localization transition}
\begin{figure}
\includegraphics[width=0.5\textwidth]{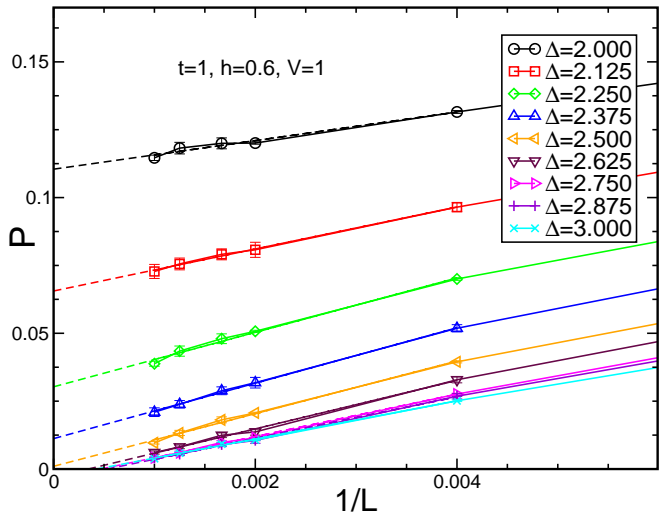}
\caption{(Color online) Finite size scaling analysis of the participation ratio.
For this value of $V$, the transition towards localization happens
at $\Delta_{\rm c} \simeq 2.5$.\label{scaling}}
\end{figure}

It is possible to analyze more precisely the transition to the localized
regime by performing a finite size analysis of the participation ratio.
We define the mean participation ratio $P = \sum_k P_k / L$ which becomes
zero when we have only localized states.
We found that the participation ratio is varying linearly with the size
of the system $L$. In Fig. \ref{scaling} we show the result of a finite
size analysis for a given value of $V$ and for different values of 
$\Delta$. We see that $P$ extrapolates to zero for values larger
than $\Delta_c \simeq 2.5$. This is in agreement with the phase diagram
obtained with a $L=1000$ size (Fig. \ref{phasediagram2}).

\section{Conclusion}

We have studied the Hatano-Nelson model in the presence of a periodic
potential.  We have determined the phase diagram of the system and shown
that it presents a variety of different localized or gapped phases,
distinguished by the topology of the eigenspectrum in the complex plane.
Interesting insulator-metal transitions can occur, driven by increasing
disorder. Finally, we have determined how the gap and participation
ratio evolve close to these transitions.

There are some interesting analogies between eigenspectra
features induced by a periodic potential, discussed here,
and those which appear in crossing over from one
to two dimensions \cite{zee98}.
We comment first that the $d=2$ model can be trivially
solved in the clean limit $\Delta=0$ with hoppings
$t_x, t_y$ and non-Hermiticity parameters $h_x,h_y$.  
In fact, the eigenvalue distribution in the
complex plane is obtained by taking 
ellipses determined by $t_y$ and $h_y$, and placing them with
centers at each
of the eigenvalues of Eq.~\ref{eq:deq0spectrum}.  In the limit
of large lattice sizes these eigenvalues become space filling in
the complex plane.  

However, just as discussed here, different topologies are possible.
If $t_y$ and $h_y$ are small compared to $t_x$ and $h_x$ (or 
{\it vice-versa}) one is overlaying small ellipses on a background of
a larger ellipse, and the eigenspectrum retains a hole in the origin.
When the hoppings in the two dimensions are comparable,
in contrast, the central hole can get filled in.
It is not surprising that a particularly close analogy 
is obtained by considering a system consisting of two chains since
such a situation describes, as with the periodic potential
considered here, a lattice with two different types of sites (two bands).
The resulting eigenvalue spectra reflect considerable similarities
with those found in this paper.  
See, for example, Fig.~2 of [\onlinecite{zee98}].

Random non-Hermitian matrices also arise in the study of quenched
Quantum Chromo-Dynamics at finite quark
density.\cite{stephanov96,halasz97}  While the structure of the 
matrices does not take the precise form of the HNM, it is
interesting that in certain cases (``Dyson index" $\beta=4$) a gap
appears in the spectrum in the complex plane, analogous to that found
here.\cite{halasz97}  
That is, the eigenvalue distribution can be driven to zero in
a region in the vicinity of the imaginary axis, not unlike the 
effect of increasing $V$ in Fig.~\ref{typical_D0}.  (In these studies the
spatial dimensionality is greater than one so that the eigenvalues
are space filling as opposed to lying along one-dimensional curves.)
Indeed, a key result\cite{stephanov96} is that
the appearance of a gap around the imaginary axis
requires a finite value of the appropriate tuning
parameter, the chemical potential $\mu$, much as our phase diagram
(Fig.~\ref{phasediagram}) requires a finite staggered
potential $V > V_c \simeq 0.7$ for
the appearance of a gap.

The original motivation for the HNM was the problem of flux line
motion in type II superconductors.  
Specifically, deliberate heavy ion irradiation
\cite{CivaleReview} produces
random tracks along which the flux lines like to sit, and
a transverse magnetic field then provides the 'non-Hermiticity'
which attempts to delocalize them.  It is also possible to
provide regularly patterned pinning centers, which would be a 
realization of the staggered $\pm V$ term in the extension of the
HNM presented here.\cite{baert95,harada96,reichhardt96}
Our phase diagram, 
Fig.~\ref{phasediagram} indicates that the amount of random
disorder $\Delta$ required to produce localized phases in
general increases with the amplitude $V$ of the patterned
potential, so that critical currents would decrease
as patterned pinning centers are established if the modulation
is at the atomic level.  It would be interesting to study the
effect of regular variation in the site energy at longer length
scales, such as would more typically be accessed in experiments
with patterned pinning.

\begin{acknowledgments}
We acknowledge support from the National Science Foundation under
award NSF-ITR-013390, the  Natural Science Foundation of Shanghai (10ZR1403400), and
ARO Award W911NF0710576 with
funds from the DARPA OLE Program.
This work was also supported by the CNRS-UC Davis EPOCAL LIA joint research grant. The authors
would like to thank B.~Holly and B.~Gr\'emaud for useful input. Z. Bai and R.T.~Scalettar
wish to thank the hospitality of Fudan University where portions
of this work were done.
\end{acknowledgments}


\end{document}